# Development of Low Cost Private Office Access Control System(OACS)

Sadeque Reza Khan

Prime University, Department of Electrical and Electronic Engineering, Dhaka-1216, Bangladesh
sadeque_008@yahoo.com

*ABSTRACT*

*Over the years, access control systems have become more and more sophisticated and several security measures have been employed to combat the menace of insecurity of lives and property. This is done by preventing unauthorized entrance into buildings through entrance doors using conventional and electronic locks, discrete access code, and biometric methods such as the finger prints, thumb prints, the iris and facial recognition. We have designed a flexible and low cost modular system based on integration of keypad, magnetic lock and a controller. PIC 16F876A which is an 8-bit Microcontroller, is used here as a main controller. An advanced simulation based compiler Flowcode V4 is used to develop the software part in this project.*

## 1. INTRODUCTION

The primary purpose of an office access control system (OACS) is to secure access-controlled zones by restricting access to zones to only those persons (or assets) who are allowed access. An access-controlled door is not a single entity but a collection of door hardware that typically includes controlled outputs, such as a door lock, door holder, door sounders etc., and supervised inputs, such as door contacts, request-to-exit inputs, motion detectors, etc [1]. An access control system has the role to verify and mediate attempts made by users to access resources in a system. An access control system maps activities and resources to legitimate users [2]. The main aim of the present work is to develop an automated low cost office access control system with the help of a microcontroller. Automation is the use of control systems and information technologies to reduce the need for human work in the production of goods and services. Here a keypad is used to restrict the access of persons and a corresponding password is provided for each individual. So correct password ensures transaction through an electromagnetically locked door whether wrong password collapses the system and generates alarm. A LCD is interfaced with the main controller to show different status when a user gives his corresponding password.

## 2. PROPOSED SYSTEM

Figure 1 shows the overall proposed system where the main controller IC obtains a user password from a keypad. If the password is correct then it will permit the access of a valid user by controlling a electromagnetic door lock. For an invalid password the system will stack and blow protection alarm.

DOI : 10.5121/ijesa.2012.2201     1

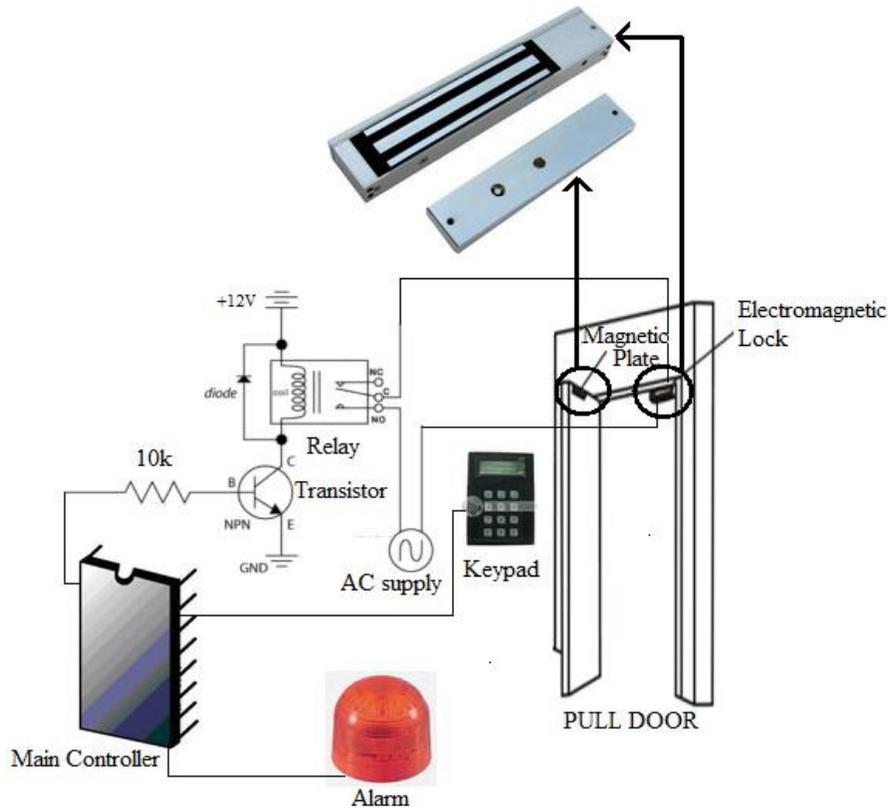

Figure 1. Proposed System.

## 3. HARDWARE DESIGN

### 3.1. Main Controller

The control module is built with the microcontroller IC. The central controller is Microchip PIC16F876A which is an 8-bit Microcontroller with up to five channels built-in A/D converter and 22 I/O pins [3]. This microcontroller IC manages overall OACS operations. It collects password from keypad, processes password and decides whether it is valid or not. One of the advantages of microcontrollers is low power consumption CMOS technology which makes them flexible for battery powered applications like this project and wide operating voltage range, for example PIC16F876A operates in a voltage range of 2-5.5 volts.

### 3.2. Display Section

The display unit used is a 16x2 (16 characters, 2 lines) alphanumeric liquid crystal display (LCD), which can be interfaced with a 4-bit or 8-bit microprocessor or microcontroller [4], [5]. LCD is widely used in recent years as compared to LEDs. This is due to the declining prices of LCD, the ability to display numbers, characters and graphics, incorporation of a refreshing controller into the LCD, their by relieving the CPU of the task of refreshing the LCD and also the ease of programming for characters and graphics[6].

### 3.3. Keypad

In this project 4x3 keypad is used. A keypad is simply an array of push buttons connected in rows and columns, so that each can be tested for closure with the minimum number of connections. The key press is scanned by bringing each X row low in sequence and detecting a Y column low to identify each key in the matrix [7],[8].

### 3.4. Electromagnetic Lock

The Magnetic lock uses an electrical current to produce a electromagnetic force. When a current is passed through the coil, the magnet lock becomes magnetized. The door will be securely bonded when the electromagnet is energized holding against the armature plate [9]. The magnetic lock is a simple locking device that consists of a magnetic lock and armature plate with no moving parts and it purely works due to the magnetic field [10], [11].

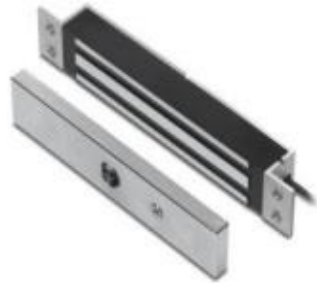

Figure 2. Magnetic Lock and Armeture Plate.

### 3.5. Alarm

An electric siren is connected in a I/O pin of PIC microcontroller through a relay to generate alarm for three consecutive wrong password entered.

## 4. LOGIC DEVELOPMENT

An advanced simulation software "Flowcode Ver. 4" is used to develop the software portion of this project. Flowcode works using flowcharts: the easiest and highest level, of programming [12], [13]. 'Flowcode' offers an easy way to program PICmicro chips. Once the flowchart is designed, on-screen, press a button and the software translates it into numerical code. For this project the component macros LCD, Keypad and LED is used in the following software which is shown in figure 3. Although we drive the magnetic lock and Alarm section with a relay but in the software we represent those two sections with two LEDs. The software Flowcode Ver. 4 provides a real time simulation which helps a lot in developing this project before going to practical hardware.

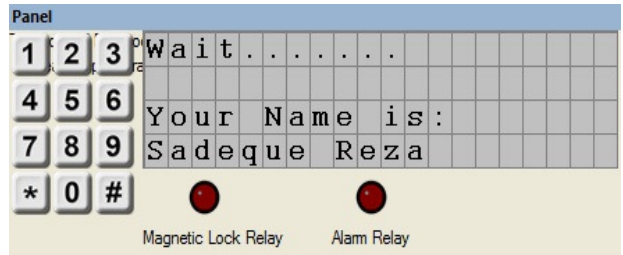

Figure 3. Simulation Output.

## 5. FLOW CHART & CALCULATION

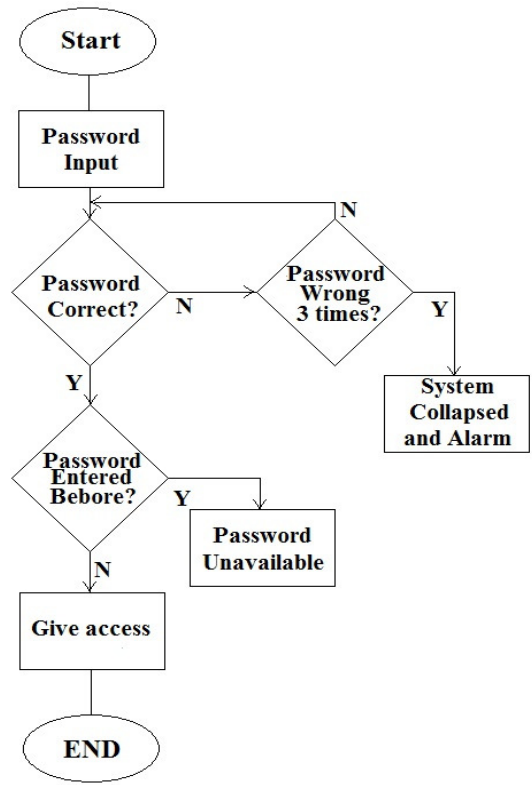

Figure 4. Logic Flowchart

In the initial state or on the free state the LCD display will show only a simple message of expecting a password from an user. If a password is inserted in the system through keypad then it will check whether the password is correct or not in its database. The user information will be inserted in the program memory of the PIC microcontroller as per private office requirements. So after inserting a password if that is not correct then the system will give chance to input password for two times more. If password is wrong for three times then the system will be collapsed automatically and an alarm will be generated. If the password matches the database then it will give access to the user by controlling a magnetic lock through a relay. The system will give access for 5 seconds and will be closed again afterwards. Again if the password is correct but it was previously entered the system will not give an access.

$$Password = A*1000+B*100+C*10+D*1$$

Here A, B, C and D are declared variables to hold the digits pressed through the keypad.

Table 1. Password Calculation Example.

| Password | Calculation | Name |
|---|---|---|
| **1234** | 1*1000+2*100+3*10+4*1 | Sadeque Reza |
| **4321** | 4*1000+3*100+2*10+1*1 | Feroz Ahmed |
| **8765** | 8*1000+7*100+6*10+5*1 | Nazmul Hossain |
| **3211** | 3*1000+2*100+1*10+1*1 | Arifa Ferdousi |

## 6. CIRCUIT DIAGRAM

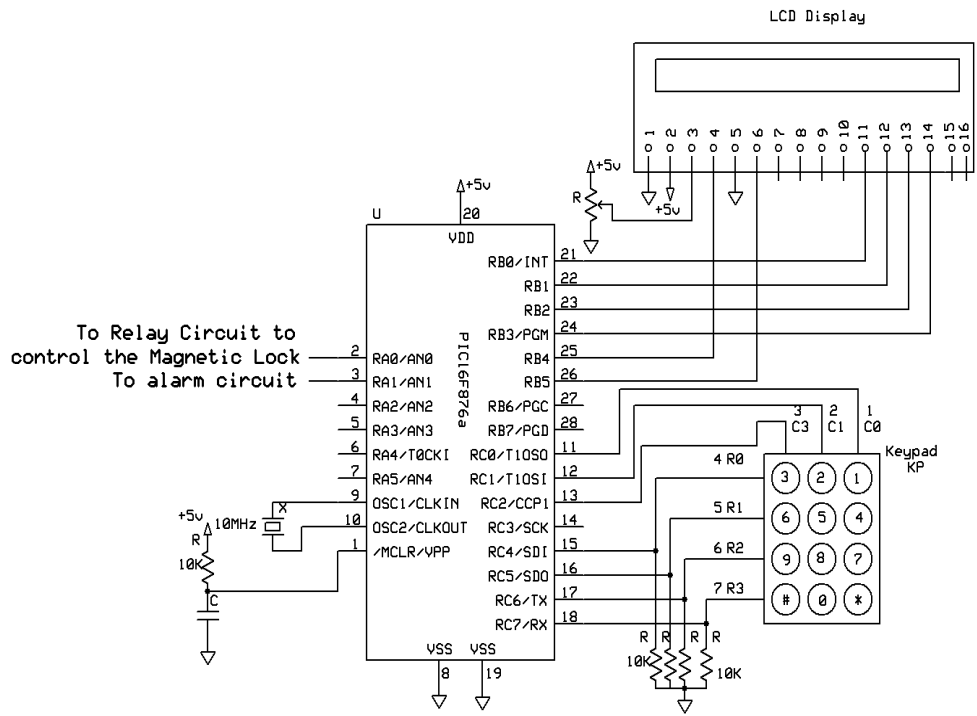

Figure 5. Implemented Circuit Diagram.

PIC16F876A is used as the main controller where LCD is connected in the PORT B and Keypad is connected in PORT C. The electromagnetic Lock and the alarm circuit are connected in PORT A. The program is inserted into the PIC by using a software and programmer called USBurn.

## 7. RESULT

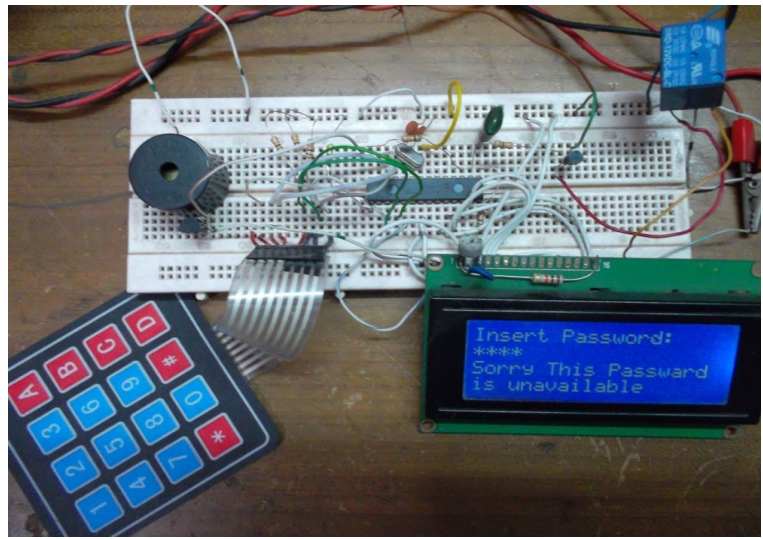

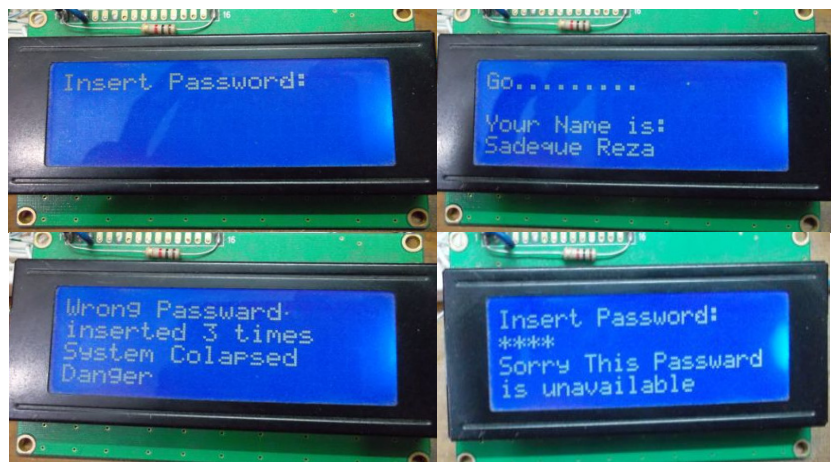

Figure 6. Overall System and Desired Outputs.

## 8. CONCLUSION

This paper has successfully presented a functional, low cost and low complexity microcontroller based office access control system. There is an option to integrate a Micro SD card or a Computer interfacing with the present system if it is required in any office system. A real-life equivalent of the prototype can be developed with minimal development costs and with relatively low operational costs for environment where high degrees of security are required like banks, military research areas, and big private investment companies. With this system we can give support up to $^{10}p_4 = 5040$ users. So this overall system can be effective not only for small private offices but also for the big houses.

## Authors

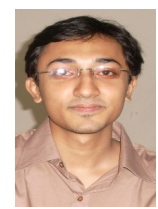

**Sadeque Reza Khan** received B.Sc. degree in Electronics and Telecommunication Engineering from University of Liberal Arts Bangladesh. Currently he is working as a lecturer in the department of Electrical and Electronic Engineering in Prime University, Bangladesh. He also works as an Advisor in an electronics goods manufacturing company named Kazi Tech. He is used to develop Microcontroller based products in this company and also gives solution for the PLC based controlling machines. His research interest includes Microelectronics, Control System Designing and Embedded System Designing.